
\input harvmac
%
\ifx\answ\bigans\else

\message{*****************************************************}
  \message{This does not look very nice in little mode.}
\message{*********************************************************************}
\fi
\voffset=-.5in

\font\blackboard=msbm10 \font\blackboards=msbm7
\font\blackboardss=msbm5
\newfam\black
\textfont\black=\blackboard
\scriptfont\black=\blackboards
\scriptscriptfont\black=\blackboardss
\def\blackb#1{{\fam\black\relax#1}}
\font\fblac=msbm9 scaled 1000
\newfam\fblack
\textfont\fblack=\fblac
\scriptfont\fblack=\blackboards
\scriptscriptfont\fblack=\blackboardss
\def\fbb#1{{\fam\fblack\relax#1}}
%

%
 
\def\BR{{\blackb R}}

%
\font\mathbold=cmmib10 \font\mathbolds=cmmib7
\font\mathboldss=cmmib5
\newfam\mbold
\textfont\mbold=\mathbold
\scriptfont\mbold=\mathbolds
\scriptscriptfont\mbold=\mathboldss

%

\Title{\vbox{\hbox{IMSc--92/31}\hbox{hepth@xxx/yymmnn}}}
{\vbox{\centerline{Correlation Functions And Multicritical Flows}
\vskip2pt \centerline {In $c<1$ String Theory}}}

\centerline{Suresh Govindarajan\foot{email: suresh@theory.tifr.res.in}}
\centerline{Theoretical Physics Group}
\centerline{ Tata Institute of Fundamental Research}
\centerline{Bombay 400 005 INDIA}
\centerline{T. Jayaraman and Varghese
John\foot{email: jayaram, john@imsc.ernet.in}}
\centerline{The Institute of Mathematical Sciences}
\centerline{C.I.T. Campus, Taramani}
\centerline{Madras 600 113, INDIA}
\bigskip
\bigskip

 We compute all string tree level correlation functions of
vertex operators in $c<1$ string theory. This is done by using the ring
structure of the theory. In order to study the multicritical
behaviour,
we calculate the correlation functions after perturbation by
physical vertex operators. We show that the $(2k-1,2)$ models can be
obtained from the $(1,2)$ model and the minimal models can be obtained
from the $(1,p)$ model by perturbing the action by appropriate
physical operators. Our results are consistent with
known results from matrix models.

\Date{September 1993}
\lref\mukhi{S. Mukhi and C. Vafa, `` Two dimensional black-hole as a
topological coset model of c=1 string theory,'', hep-th/9301083.}
\lref\SY {S. Yamaguchi, ``Correlation Functions of $(2k-1,2)$ Minimal
Matter Coupled to 2D Quantum Gravity, preprint STUPP-92-132=hep-th/9209112.}
\lref\SH {S. Hosono, Phys. Lett. {\bf B285} (1992), 35.}
\lref\ST {N. Sakai and Y. Tanii, Prog Theor.Phys {\bf 86} (1991), 547;
Int. Jour. Mod. Phys. {\bf A6}(1991), 2743.}
\lref\hyper{M. Abramowitz and I. A. Segun, {\it Handbook of
Mathematical Functions}, Dover(1964).
Z. X. Wang and D. R. Guo, {\it Special Functions}, World
Scientific(1989).
R. P. Agarwal, {\it Generalised Hypergeometric Series} Asia
Publishing House (London) 1963.}
\lref\DDK {F.David,Mod.Phys.Lett{\bf A3} (1988), 1651; J.Distler and H.Kawai,
\npb{\bf 321} (1989), 509. }
\lref\BLNW{M. Bershadsky, W. Lerche, D. Nemeschansky and N.P.
Warner, ``Extended N=2 Superconformal Structure of Gravity and
W-Gravity Coupled to
Matter,''Preprint hepth@xxx/9211040(November 1992).}
\lref\SAR{H. Kanno and H. Sarmadi, Talk presented by H. Sarmadi at the
Summer Workshop on Strings at ICTP, Trieste(July 1992).}
\lref\SARa{H. Kanno and H. Sarmadi, ``BRST cohomology ring in 2D gravity
coupled to minimal models,'' ICTP preprint IC/92/150 =
hepth@xxx/9207078.}
\lref\swapna{D. Ghoshal and S. Mahapatra, \mpl{\bf 8}(1993), 197.}
\lref\DOY{M. Doyle, ``Dilaton Contact Terms in the Bosonic and Heterotic
Strings,'' Princeton preprint PUPT-1296(1992) =
hepth@xxx/9201076.}
\lref\BT{R. Bott and L. W. Tu, {\it Differential Forms in Algebraic
Topology}, Springer-Verlag(1982).}
\lref\KACH{S. Kachru,  Mod. Phys. Lett. {\bf A7} (1992), 1419.}
\lref\ITOH{K. Itoh, ``SL(2,R) current algebra and spectrum in two
dimensional gravity,'' Texas A\&M preprint CTP-TAMU-42/91(1991).}
\lref\AGG{L. Alvarez-Gaum\'e and C. Gomez, ``Topics in Liouville Theory,''
Lectures at Trieste Spring School, CERN preprint CERN-TH.6175/91, (1991).}
\lref\GRO{U. H. Danielsson and D. Gross, \npb{\bf 366}(1991), 3.}
\lref\FKa{P. Di Francesco and D. Kutasov, \npb{\bf 342}(1990), 589.}
\lref\FKb{P. Di Francesco and D. Kutasov,\npb{\bf 375}(1992), 119.}
\lref\DOT{V. Dotsenko, \mpl{\bf 6}(1991), 3601.}
\lref\GF{G. Felder, \npb317 (1989), 215.}
\lref\LZ{B. Lian and G. Zuckerman, \plb {\bf 254} (1991), 417;  \cmp{\bf 145}
(1992), 561.}
\lref\IMM{C. Imbimbo, S. Mahapatra and S. Mukhi,``Construction of Physical
States of Non-trivial ghost number in $c<1$ String Theory,'' \npb375(1992),
399.}
\lref\GLI{M. Goulian and B. Li, \prl{\bf 66} (1991), 2051.}
\lref\BMP{P. Bouwknegt, J. McCarthy and K. Pilch, \cmp145 (1992), 541.}
\lref\BMPa{P. Bouwknegt, J. McCarthy and K. Pilch, ``Fock Space Resolutions
of the Virasoro highest weight modules with $c\leq1$,'' Lett. Math. Phys.
{\bf 23} (1991), 3601.}
\lref\KIT{Y. Kitazawa, \plb {\bf 265}(1991), 262.}
\lref\KITa{Y. Kitazawa, ``Puncture Equation in $c=1$ Liouville
gravity,'' TIT preprint TIT(1991).}
\lref\KM{M. Kato and S. Matsuda in {\it Advanced Studies in Pure Mathematics},
Vol. 16, ed. H. Morikawa(1988), 205.}
\lref\POL{A. M. Polyakov, \mpl{\bf 6}(1991), 635-644.}
\lref\FF{B. Feigin and D. Fuchs, ``Representations of the Virasoro algebra,''
in {\it Seminar on Supermanifolds} No.5, ed. D. Leites(1988), Univ. of
Stockholm Report No. 25.}
\lref\JD{J. Distler, \npb{\bf 342} (1990), 523.}
\lref\DN{J. Distler and P. Nelson, ``New discrete states of strings
near a black hole,'' Penn preprint UPR-0462T(1991).}
\lref\KAL{S. Kalyana Rama, ``New special operators in W-gravity theories,''
Tata preprint TIFR/TH/91-41(1991).}
\lref\SEI{N. Seiberg,``Notes on Quantum Liouville Theory and Quantum Gravity,
,'' Prog. of Theo. Phys., {\bf 102}(1990), 319.}
\lref\POLC{J.Polchinski, ``Remarks on the Liouville field theory,'' Texas
preprint UTTG-19-90, in Proceedings of Strings '90.}
\lref\BER{M. Bershadsky and I. Klebanov, \prl{\bf 65} (1990), 3088;
\npb{\bf 360} (1991), 559.}
\lref\KAED{K. Aoki and E. D'Hoker,  Mod. Phys. Lett.{\bf A7}
235(1992).}
\lref\descent{S. Govindarajan, T. Jayaraman, V. John and P. Majumdar,
Mod. Phys. Lett. {\bf A7}, (1992) 1063-1077.}
\lref\DOTa{Vl. M. Dotsenko, `` Remarks on the physical states and the
chiral algebra of 2d gravity coupled to $c\leq1$ matter,''
PAR-LPTHE 92-4(1992) = hepth@xxx/9201077;
Mod. Phys. Lett.{\bf A7} (1992), 2505.}
\lref\EW{E. Witten, \npb{\bf 373} (1992), 187.}
\lref\KMS{D. Kutasov, E. Martinec and N. Seiberg, \plb{\bf 276}(1992), 437.}
\lref\KLEB{I. R. Klebanov, ``Ward Identities in 2d gravity,''
Mod. Phys. Lett. {\bf A7}, (1992)723,}
\lref\CDK{N. Chair, V. Dobrev and H. Kanno, \plb{\bf 283} (1992), 194.}
\lref\EWBZ{E. Witten and B. Zweibach, ``Algebraic structures and
Differential Geometry in 2d string theory,'' \npb{\bf 377}(1992), 55.}
\lref\JDTG{J. Distler, Nucl. Phys. {\bf B342} (1990),  523.}%
\lref\ring{S. Govindarajan, T. Jayaraman and V. John, ``Chiral Rings
and Physical States in $c<1$ String Theory,'' \npb{\bf 402} (1993), 118.}
\lref\corr{S. Govindarajan, T. Jayaraman and V. John, ``Genus Zero
Correlation Functions in $c<1$ String Theory,'' Phys. Rev. {\bf D} {\bf 48}
(1993), 839.}
\lref\bershad{M. Bershadsky and D. Kutasov,
\npb {\bf 382} (1992), 213.}
\lref\matrixa{D. Gross and A. Migdal, \prl.{\bf 64} (1990), 127.}
\lref\matrixb{R. Dijkgraaf and E. Witten, \npb {\bf 342} (1990), 486.}
\lref\DF{ V. Dotsenko and Fateev, \npb {\bf 251}, (1985), 691.}
\def\cmp{Comm. Math. Phys.}
\def\plb{Phys. Lett. {\bf B}}
\def\prl{Phys. Rev. Lett.}
\def\mpl{Mod. Phys. Lett. {\bf A}}
\def\npb{Nucl. Phys. {\bf B}}

\def\lfr#1#2{\textstyle{#1 \over #2 }}
\def\al{{\alpha}}
\def\pa{{\partial}}

\def\gam#1#2{\Gamma({#1}_{#2})}

\newsec{Introduction}
  The complete solvability of the matrix models has provided string theory
with  the perfect scenario in which to study various non-perturbative
and stringy effects. However in order to translate this success into the
context of a generic string theory for which a matrix model formulation
may not exist, one needs to understand the matrix models in the continuum
language. Hence, this has led to an extensive study of the continuum
formulation of matrix models, i.e., $c<1$ matter coupled to Liouville gravity
 \refs{\DDK\POL\GLI\BER\ST\DOT\KIT\swapna\KAED \FKb \AGG
\descent-\SH}.
   The algebraic structure of the $c=1$ string theory i.e, the presence
of the ground ring has been translated for the $c<1$ case in
\refs{\KMS \CDK \ring -\SARa}. In this paper, we continue the progress
of earlier work and obtain all the correlation functions on the sphere
for all minimal models coupled to Liouville gravity. We also reproduce
the multicritical behaviour seen in the matrix models.

The analysis of non-critical strings in the conformal gauge
(using the DDK ansatz\DDK)
 was able to predict the susceptibility exponent and the
 area scaling exponents of gravitationally dressed primary fields in the
 theory. Subsequently, the three point functions of vertex operators in the
 theory were obtained by doing the zero mode integration over the Liouville
field and then evaluating the three point functions using an analytic
 continuation in the number of cosmological constant operator insertions
\GLI\DOT\KIT . To obtain all the
matrix model three point functions the vertex operator states with momenta
outside the conformal grid had to be included.

The cohomology analysis required to obtain the physical states of the theory
was done by Lian and Zuckerman\LZ\ and  Bouwknegt et al \BMP.
Their analysis showed that there was an infinite number of physical states of
non-trivial ghost number. Half of these states (with $\beta < \beta_0
$)
had scaling exponents which matched those seen in the matrix models.
In \descent, it was shown that the states of non-trivial ghost number were
related to pure vertex operator states (DK states) using descent equations
arising from the double cohomology used to obtain the physical states. This
explains the presence of states outside the Kac's table.

 For $c=1$ matter coupled to Liouville gravity, it was shown by Witten\EW\
that there exists a ring of operators of zero ghost number with a
multiplication rule given by the usual OPE. In the context of $c<1$ matter
coupled to Liouville gravity, Kutasov et. al. obtained a finite ring
of ghost number zero operators\KMS. This ring was extended in \SARa\ to an
infinite ring of operators at all ghost numbers. However, in \KMS\SARa, the
matter sector was represented by the Verma module of the minimal models.
If the matter sector is represented by its free field realisation \`a la
Feigin-Fuchs, the ring of ghost number zero operators is infinite as shown
in \ring\foot{It was also suggested in \CDK \DOTa  \ITOH that a
$SO(2, {{\fbb C}})$
rotation of the fields was enough to relate the $c<1$ and the $c=1$ models
coupled to gravity. As was pointed out in \ring, this is not
sufficient since the role of the Felder BRST operator must be taken into
account. This crucially differentiates the properties of a rotated $c=1$
model and the $c<1$ model coupled to Liouville gravity.}. The states belonging
to the edge of the Kac table whose Liouville scaling exponents are observed
in the matrix models do not occur in the formalism of \KMS\SARa. These states
do not decouple in correlation functions\KIT\DOT\ and hence have to be
included. When the free field realisation of the minimal models is used,
the edge states have to be included. Hence, the infinite ring structure
of \ring\ has all the exponents seen in the matrix models. In
addition, the ring is identical to that seen in the case of topological
minimal matter coupled to topological gravity.

    Having obtained the ring structure of these models, it is natural to
see if it is possible to obtain all the correlation functions in the theory
using the ring. In this paper, we will discuss in detail how the ring structure
can be used to obtain all the genus zero
correlation functions in the $c<1$ models\corr.
The approach we adopt here is quite general
and can be applied to other problems which have the ring structure
such as the non-critical superstring, the W-string, and the $SL(2,\BR)/U(1)$
(black-hole) coset model. We will also show that it is
possible to study the behaviour of the correlation functions under perturbation
 by physical operators in the theory and determine their behaviour as function
of the coupling constants; and that it is possible to obtain multicritical
behaviour in these models which generalises the results of Distler\JDTG.

The paper is organised as follows. In section 2, we describe the
vertex operator states and the ring of a generic $c<1$ model coupled to
Liouville gravity. In section 3, we
compute the correlation functions for the $(2k-1,2)$ models.
In section 4, we study the $(p+1,p)$ minimal models coupled to gravity.
We also compare the results with known KdV/matrix model results.
Finally, in section 5, we reproduce the multicritical behaviour seen in
the matrix models using the correlation functions computed in the
earlier sections. This completes the proof of equivalence of the
matrix models and the continuum formulation at the string tree level.

\newsec{Physical States And Rings}
    Let us consider the (p',p) model coupled to gravity.
We consider two scalars $X$(for matter) and $\phi$(for the Liouville
mode) with background charges $\al_0$ and $\beta_0$ respectively at infinity.
 The corresponding energy-momentum tensors are given by
\eqn\estress{\eqalign{
T^M &= -\lfr14\pa X\pa X + i \al_0 \pa^2 X\quad,\cr
T^L &= -\lfr14\pa\phi\pa\phi + i \beta_0 \pa^2\phi\quad,
}}
with central charges $c_M=1-24\al_0^2$ and $c_L=1-24\beta_0^2$. For
the
$(p',p)$ minimal models
$$
\al_0^2=\lfr{(p'-p)}{4pp'}~ {\rm and}~\beta_0^2=
-\lfr{(p+p')^2}{4p(p+1)}\quad.
$$
 The vertex operators $e^{i\al X}$ and $e^{i\beta\phi}$ have conformal weights
$\al(\al-2\al_0)$ and $\beta(\beta-2\beta_0)$ respectively.
The gravitationally dressed vertex operators are of the form
$exp(i\alpha X + i\beta \phi)$    with conformal dimensions given by
$ \alpha(\alpha-2 \alpha_0)+\beta (\beta- 2 \beta_0)=1$ .
The usual matter screening operators are given by :
$$Q_+ =\Delta  (1+ {p' \over p})\int e^{i \alpha_+ X} ,\quad
  Q_- =\Delta  (1+ {p\over p'})\int e^{i \alpha_- X}
$$
where
 $$
\al_+=\sqrt{\lfr{p'}p}~{\rm and}~\al_-=-\sqrt{\lfr p{p'}}\quad.
$$
and  $ \Delta (x)= {\displaystyle {\Gamma(x)\over \Gamma(1-x)}}$.
 (note the measure factor
${\displaystyle {d^2 z \over 2\pi} }$ is implicit. The normalisation of
the operators is explained later.)

Felder\GF, has shown that the screening operators have a BRST like
action which enables one to truncate the large space of Virasoro primaries
of a scalar field to the finite set of primaries seen in the minimal
models. Before coupling to gravity, the two screening operators play
identical roles. However, on coupling to gravity this {\it duality is
broken}\corr. The reason for this is as follows. Suppose we choose to
use the screening operator $Q_+$
to function as the BRST operator\foot{This will be referred to as the
$Q_+$ resolution. The choice of $Q_-$ as the Felder BRST gives the
$Q_-$ resolution.},
then $e^{i\al_- X}$ becomes a physical operator. Hence, one can no
longer use $Q_-$ as a screening operator. Further, the spectrum of
physical states in the two resolutions are seen to be different\corr.
This is not surprising. In the matrix model, the $(p',p)$ model can be
obtained in two different ways -- as a the $p$-th critical point of
the $(p'-1)$ matrix model or the $p'$-th critical point of the $(p-1)$
matrix model. The spectrum of the two resolutions agree with those
seen in the two different matrix models. Later, we shall explicitly
demonstrate this difference.

It has been shown\descent\ring, that using the descent equations
arising from the double cohomology of the string and Felder BRST
operators that the physical operators with
Liouville charge $\beta<\beta_0$ exist at two ghost numbers:

$--$ the ghost number $1$ operators correspond
to pure vertex operators of the form  $c{\bar c}exp(i\alpha X+ i\beta
\phi)$. These are the DK states\DOT\KIT\descent. The allowed values of
matter momenta take values both outside and inside the conformal
grid(Kac Table).

$--$ the ghost number $0$ operators which form the chiral ring\ring.

\subsec{DK States}

The DK  operators  for $(p',p)$ models in the $Q_-$ resolution
are given by
\eqn\verta{
 V^\al_ n =\Delta(n + \lfr{\al+1}{p})
exp {\displaystyle {[p(n-1)+\alpha + 1-p']\phi
+[p(n-1)+\alpha + 1 + p']iX
\over{2 \sqrt {p'p}}}}}
 where  $\alpha =0,\ldots,(p-2)$.
As was explained in \descent, these include all vertex operator states
with matter momenta corresponding to values occuring to the left of
the central Fock tower $F_{m',m}$ in the Felder
complex(with $0<m<p,~0<m'<p'$) . The Fock
tower labels are $F_{\pm m'+2jp',m}$ for $j=1,2,\ldots$. The states
corresponding to the {\it edge of the Kac Table} with labels $(jp',m)$ are
included since they do not decouple after coupling to gravity\KIT.
However, in the $Q_-$ resolution, the other edge states with labels
$(m',jp)$ are not included since they are (Felder)BRST exact. So
unlike the states inside the Kac Table, the states which belong to the
edge of the Kac table come with the resolution. In the $Q_+$
resolution, edge states with labels $(m',jp)$ are no longer
(Felder)BRST exact. However, the other set of edges are now trivial.
This leads to the following set of DK states in the $Q_+$ resolution.
\eqn\vertb{
 V^a_ n = \Delta(n + \lfr{\al+1}{p'})
{exp {\displaystyle {[p'(n-1)+\alpha +(1-p)]\phi
+[-p'(n-1)-\alpha-(1+p)]iX \over{2 \sqrt {p'p}} }}}}
where  $\alpha =0,\ldots,(p'-2)$.

 All the operators
 are normalised  by a momentum-dependant ratio of $\Gamma$ functions of the
form $\Delta (n+ {(\alpha +1)\over p})$ for the operators appearing in the
$Q_-$ resolution and by
$\Delta (n+ {{\alpha +1}\over{p'}})$ for the operators appearing in the
$Q_+$ resolution. (Notice that this normalisation is finite for all
the allowed values of $\alpha$ .) This normalisation removes all the external
leg factors corresponding to the operators in the scattering amplitudes,
and provides the correct normalisation to compare with the matrix models.

    In the calculation of the correlation functions, we will use only physical
operators that appear in one resolution.
This is essential to obtain the right combinatorial
factors required to reproduce the matrix model results.  Another
benefit of not mixing resolutions(as we shall see soon) is that the
ring structure is identical to that seen in topological matter coupled
to topological gravity.
 Thus it seems that on coupling to gravity the usual
minimal model duality is lost.  Furthermore, only one screening operator
can be inserted  to obtain charge conservation in the matter sector.
Another interesting point to note here, is that the DK states on being
Lorentz boosted to a compact $c=1$ model, correpond to tachyons at a
particular radius. However, the screening operator corresponds to a
tachyon of a different radius indicating that this theory is not a
Lorentz boosted $c=1$ theory.

\subsec{Ring Structure}

Apart from the vertex operator states there are an infinite set of operators of
zero ghost-number in the double cohomology, which form a ring under
multiplication.
  The chiral ring is generated by two elements\KMS,\ring
\eqn\egena{\eqalign{
x&=\left (b_{-2}c_1 + t(L_{-1}^{L} - L_{-1}^{M})\right )~~
e^{{1\over{2t}}(-iX+\phi)}
\cr
y&=\left (b_{-2}c_1 + \lfr1t(L_{-1}^{L} - L_{-1}^{M})\right )~~
e^{{t\over2}(iX+\phi)}
}}
where $t=\sqrt {p\over p'}$.
Again, just as in the case of the DK states, the ring obtained is
dependent on the choice of resolution. For the $Q_-$ resolution, the
chiral ring is given by
\eqn\erela{
\{ x^m~y^n~|~ x^{(p'-2)}=0 \}\quad.
}
It can be seen that $x^{(p'-2)}$ is $Q_F$-exact. However, this is no
longer true if $Q_+$ is used as the Felder BRST operator. Now,
$y^{p-2}$ is $Q_F$ exact. So the chiral ring in the $Q_+$ resolution is
given by
\eqn\erelb{
\{ x^m~y^n~|~y^{p-2}=0 \}\quad.
}
The ring structure seen here is identical to that seen in the case of
topological matter coupled to topological gravity. In the $Q_-$
resolution, the ring generated by $x$ is identical to that of
topological matter with $y$ corresponding to the infinite ring of
topological gravity. This is suggestive of an underlying twisted $N=2$
supersymmetry in this theory\foot{See a recent paper by Bershadsky,
et. al.\BLNW\ where this has been explicitly demonstrated. Also, see
\mukhi.}

In \descent, an equivalence relation had
been obtained by demanding that both resolutions be identical. There,
the role of the edge states was unclear and the decoupling of the so
called `wrong-edge' was not obvious. However, by staying in one
resolution as in this paper, these questions are easily answered
as we have shown. The wrong-edge states are Felder-exact and hence do
not belong to the set of physical states. This explains the relations
\erela \erelb\ as opposed to the equivalence relation obtained in
\descent.

  So far we have  introduced  only the holomorphic part,
however keeping in mind the fact that the Liouville boson is non-compact
we require that the holomorphic-antiholomorphic Liouville momenta are equal.
This leaves us with two elements:
\eqn\ering{\eqalign{
a_+ =& x \bar x \cr
a_- =& y \overline y \quad,
}}
with the appropriate ring relation according to the resolution chosen.
These ring elements through their action on the DK states provide relations
between correlation functions using which we find some operator identities.
These operator identities we will show (in the next section) are enough to
determine the correlation functions completely.

\newsec{Correlation Functions in $(2k-1,2)$ Models}
 In this section we will calculate the n-point correlation functions
in the $(2k-1,2)$ models using the action of the rings on the Vertex operator
states.
 The partition function is given by:
\eqn\pfunc{
{\cal Z} = \int {\cal D} \phi~ {\cal D} X~ {\cal D}b{\cal D}c ~
\displaystyle {e^{-S_M - S_L -S_{gh}+ \mu \int V_0 }}
}
where the matter and Liouville theories are defined in terms of free bosons
with charges at infinity given by:
$$
\al_0^2=\lfr{(2k-3)^2}{8(2k-1)}~ {\rm and}~\beta_0^2=
-\lfr{(2k+1)^2}{8(2k-1)}\quad.
$$

 Consider the correlation function
\eqn\eqcorr{
\langle \langle \prod_{i=1}^{L} \int V_{n_i} \rangle \rangle =
\mu^S \Gamma(-S) \langle c{\bar c}V_{n_1}(0)c{\bar c}V_{n_2}(1)
c{\bar c}V_{n_3}(\infty)
\prod_{i=4}^{L} \int V_{n_i} (\int V_0)^S
{1\over{R!}}
(Q_-)^R \rangle ,
}
where the average is with respect to the action defined in \pfunc\
The RHS is obtained after doing the zero mode integration over the
Liouville field as was done in \GLI.  The values of R,S are determined by the
requirement of charge conservation, in the Liouville and matter sectors.
For the moment we will consider only the correlation functions that require
a positive integer number of screening operators. After determining the
dependence of the amplitudes on the screening operators we will analytically
continue the results to evaluate the correlation functions that require
insertion of negative integer or fractional number of screening operators
for charge conservation.
 The vertex operators in  the $Q_-$ resolution in this theory are given
by :
\eqn\vert{
V_n = \Delta(n+{1\over2})~~ exp {{(n-k)\phi + (n+k-1)iX} \over
{\sqrt{2(2k-1)}}}\quad,
}
where $n=0,1,\ldots$.
These models are simpler than the generic $(p+1,p)$ models in the sense that
there is only one allowed value of $\alpha$ and hence the algebra required to
obtain the correlation functions is straightforward and transparent.
Note that the ``physical'' screening operator $Q_+$ is given by $V_k$.
The charge conservation relations for the correlation functions given in
\eqcorr\ are given by
\eqn\cons{\eqalign{
 \sum_{i=1}^L &n_i - kL - Sk = - (2k+1)\quad, \cr
 \sum_{i=1}^L &n_i + kL -L+ (k-1)S + (2k-1)R = 2k-3\quad.\cr
}}

The ring elements are generated by a single generator $a_-$ \ring\ , where
\eqn\eqring{
a_- = -|bc + {1\over2}\sqrt{{(2k-1)}\over{2}}\partial(\phi - iX)|^2
{}~~exp{{(\phi + iX)}\over {\sqrt{2(2k-1)}}}\quad.
}
(Since the Liouville field is treated as a non-compact boson with a charge at
infinity the Liouville momenta in the holomorphic and anti-holomorphic sectors
is taken to be the same.) The ring elements are $(a_-)^n$. The action of the
ring element on the DK states is given by \KMS \KACH \bershad
\eqn\action{
\lim_{z\rightarrow w}a_-(z) c{\bar c}V_n(w) \sim c{\bar c}V_{n+1}(w),
\quad a_-(z) c{\bar c}e^{i\alpha_-X}(w)\sim 0 \quad.
}
Now consider a charge conserving correlation function with one $a_-$
and DK states

\eqn\ercorr{
F(w,{\bar w}) \equiv \langle a_-(w) c{\bar c}V_{n_1}(0) c{\bar c}V_{n_2}(1)
c{\bar c}V_{n_3}(\infty) \prod_{i=4}^{N} \int
V_{n_i} {1\over{R!}}(Q_-)^R\rangle \quad.
}

One can check that $\partial_w F=\partial_{\bar w} F=0$ using
$\partial_w a_- =\{Q_B, b_{-1}a_-\}$ and then deforming the contour of
$Q_B$. This implies that $F(w,{\bar w})$ is a constant independent of
$w$ and ${\bar w}$. Equating $F(0)$ with $F(1)$ and using
\action , we obtain
\eqn\reln{\eqalign{
&\langle  c{\bar c} V_{n_1+1}(0) c{\bar c} V_{n_2}(1)
c{\bar c} V_{n_3}(\infty) \prod_{i=4}^{N} \int
 V_{n_i} {1\over{R!}} (Q_-)^R\rangle
\cr
=&\langle  c{\bar c} V_{n_1}(0) c{\bar c} V_{n_2+1}(1)
c{\bar c} V_{n_3}(\infty) \prod_{i=4}^{N}  \int
 V_{n_i}{1\over{R!}}(Q_-)^R\rangle\quad.
}}
This gives us the following recursion relation(similar to the one in
\JDTG )
\eqn\relna{
V_{n+1}(z)V_m(w)
= V_{n}(z)V_{m+1}(w)  \quad.
}
In particular, $V_nV_0=V_{n-k}V_k$. This is sufficient to convert all
correlation functions(with positive integer number of Liouville
screening)  to ones which are of the Dotsenko-Fateev type\DF\
making them computable.
We can now use the recursion relation, to obtain
\eqn\kcorrg{\eqalign{
&\langle \langle  \prod_{i=1}^{N} \int  V_{n_i} \rangle \rangle =
{\mu ^S }~~\Gamma (-S)~~
\langle c{\bar c}V_{0}(0)c{\bar c}V_{0}(1)~c{\bar c}
 V_{k-1}(\infty)
(\int V_2^0)^{X} {1\over{R!}} (Q_-)^R \rangle \quad\cr
&={\mu ^S }~ ~\Gamma (-S)~ ~ {\cal N}~ ~
\prod _{i=1}^X \prod _{j=1}^R \int d^2z_i d^2w_j
 {| z_i| }^{2(2k-1)}
{| 1-z_i| }^{2(2k-1)}
\prod_{i<j} {| z_i-z_j| }^{2(2k-1)}\cr
&\quad \times {| w_i| }^{-2(k-1)\rho }
{| 1-w_i| }^{-2(k-1)\rho  } \prod_{i<j}{| w_i -w_j| }^{4\rho }
\prod_{i,j}{| z_i-w_j| }^{-4}\cr
&={\mu ^S }~ ~\Gamma (-S)~ ~ {\cal N}~ ~ ({ X!})~ ~(\pi)^{X+R}~ ~ (\rho)^{4RX}
{}~ ~\Delta (1-\rho')^X~ ~ \Delta (1-\rho)^R\cr
\times &\prod_{i=1}^X \Delta (i\rho' -R) \prod_{i=1}^R\Delta (i\rho)\cr
\times &\prod_{i=0}^{X-1} \Delta (k-R+i\rho')
 \Delta (k-R+i\rho') \Delta (R-2k+1)-(X-1+i)\rho')\cr
\times &\prod_{i=0}^{R-1} \Delta (1-{(k-1)\rho}+i\rho)
\Delta (1-(k-1)\rho +i\rho)
\Delta (2X-1-{2(k-1)\rho}-(R-1+i)\rho) \cr
 }}
where $\Delta(x) ={\Gamma(x)\over\Gamma(1-x)}$ ; $X =L+S-3$  ;
$\rho={2\over 2k-1}$ ; $\rho'={2k-1\over 2}$ and\quad
${\cal N }$ is the
normalisation factor associated with the operators.

This correlator was evaluated using formula (B.10) of
\DF. This gives
$$
\langle \langle \prod_{i=1}^{L} \int  V_{n_i} \rangle \rangle
= \rho~\mu^S~ \Gamma(-S) ~(L+S-3)! {{\Gamma(1)}\over{\Gamma(0)}}
=\rho~\mu^S~{{\Gamma(L+S-2)}\over{\Gamma{(S+1)}}}\quad,
$$
where $\rho = {2\over{(2k-1)}}$.
and using \cons
$$S= (2-L) + {1\over k}{(\sum_{i=1}^L n_i) +{1\over k}}$$
$$2R=2 (\sum_{i=1}^L n_i)- (L+S-4)$$.

The cases of non-integer $S$ and $R$ are obtained by analytically
continuing the above result to non-integer values. Hence, one obtains
\eqn\result{
\langle \langle \prod_{i=1}^{L} \int V_{n_i} \rangle \rangle =
\rho~\mu^S~{{\Gamma({{(\sum_in_i) +1}\over {k}}})\over {\Gamma(S+1)}}\quad,
}
This answer is the same as the matrix model results \matrixa\matrixb \JDTG.
The use of the ring to obtain operator relations has simplified the
computations enormously. These correlation functions have been
subsequently calculated using different techniques in \SY. However, there
the role of the Felder BRST charge, especially in the choice of resolution
is not clear.

\newsec{Correlation Functions In $(p+1,p)$ Models}
      In this section we will consider the $(p+1,p)$ models or the unitary
minimal models coupled to gravity\foot{We have restricted our
discussion to the unitary minimal model for simplicity. The extension
to the arbitrary $(p',p)$ model is straightforward.}.
Here the number of allowed values of
$\alpha$ as introduced in \verta\ are $\alpha = 1...p-2$.( Notice that
the number of allowed values of $\alpha$ is the same as the number of primary
fields in the corresponding topological sigma models coupled to gravity.)
The DK states in the $Q_-$ resolution are
$$
V_n^\alpha = \Delta(n + {{(\alpha+1)}\over p})exp{{[p(n-2)+\alpha]\phi
+ [pn+\alpha+2]iX } \over {2\sqrt{p(p+1)}}}
$$
where $\alpha=0,\ldots,(p-2)$.

The ring elements for these models are generated by\KMS
\eqn\ringp{\eqalign{
a_+&=  -|bc + {1\over2}\sqrt{{p}\over{p+1}}\partial(\phi + iX)|^2
{}~~exp{{(p+1)(\phi - iX)}\over
{2\sqrt{p(p+1)}}},\cr
a_- &= -|bc + {1\over2}\sqrt{{p+1}\over{p}}\partial(\phi - iX)|^2
{}~~ exp{{p(\phi + iX)}\over {2\sqrt{p(p+1)}}}\quad.}}
The ring elements in the $Q_-$ resolution are \ring
\eqn\ringe{
(a_-)^n,~~ a_+(a_-)^n,~~ \ldots,~~(a_+)^{p-2}(a_-)^n\quad.
}
with the relation $a_+^{p-1}=0$.
The action of the ring on DK states is\KMS ,\KACH,\bershad\
\eqn\eringb{\eqalign{
&\lim_{w\rightarrow z}a_-(w)c{\bar c}V_n^\alpha(z) \sim c{\bar c}
V_{n+1}^\alpha(z)\quad,\cr
&\lim_{w\rightarrow z}a_+(w)c{\bar c}V_n^\alpha(z)\int V_m^\beta(t)
\sim c{\bar c} V_{n+m-1}^{\alpha+\beta +1}(z)\quad,\cr
}}
where we have normalised all the DK states by their appropriate
leg-factors.
The n-point functions can be
obtained as before (after doing the zero mode integration) we have
\eqn\pcorra{
\langle \langle  \prod_{i=1}^{N} \int  V_{n_i}^{\alpha_i} \rangle \rangle =
{\mu^S }\Gamma(-S)
\langle c{\bar c}V_{n1}^{\alpha_1}(0)c{\bar c}V_{n2}^{\alpha_2}(1)~c{\bar c}
V_{n3}^{\alpha_3}(\infty) \prod_{i=3}^{N} \int  V_{n_i}^{\alpha_i}
(\int V_0^0)^S{1\over{R!}} (Q_-)^R \rangle \quad }
       The charge conservation relations are
\eqn\conspb{\eqalign{
p\sum_{i=1}^N n_i&+ \sum_{i=1}^N\alpha_i- 2pL - 2pS = - (4p+2)\quad,{\rm
and}\cr
p\sum_{i=1}^N n_i&+ \sum_{i=1}^N\alpha_i+ 2L + 2S-2pR = 2 \quad.\cr
}}
Now we can use the ring elements to obtain operator relations.
If we insert the ring element $a_-$ into a correlation function and use the
the fact that the correlation function is independent of the position of the
ring element then we have
\eqn\rcorr{
F(w,{\bar w}) \equiv \langle a_-(w)
c{\bar c}V_{n1}^{\alpha_1}(0)c{\bar c}V_{n2}^{\alpha_2}(1)~c{\bar c}
V_{n3}^{\alpha_3}(\infty) \prod_{i=3}^{N} \int  V_{n_i}^{\alpha_i}
(\int V_0^0)^S{1\over{R!}} (Q_-)^R \rangle
}
\eqn\ind{
\lim_{w\rightarrow 0}F(w,{\bar w}) = \lim_{w\rightarrow 1}F(w ,{\bar w})
}

which implies that
\eqn\relnp{\eqalign{
&\langle c{\bar c}V_{n1+1}^{\alpha_1}(0)c{\bar c}V_{n2}^{\alpha_2}(1)~c{\bar c}
V_{n3}^{\alpha_3}(\infty) \prod_{i=3}^{N} \int  V_{n_i}^{\alpha_i}
(\int V_0^0)^S{1\over{R!}} (Q_-)^R \rangle
\cr
=&\langle c{\bar c}V_{n1}^{\alpha_1}(0)c{\bar c}V_{n2+1}^{\alpha_2}(1)~c{\bar
c}
V_{n3}^{\alpha_3}(\infty) \prod_{i=3}^{N} \int  V_{n_i}^{\alpha_i}
(\int V_0^0)^S{1\over{R!}} (Q_-)^R \rangle
 \quad.
}}
This gives us the following operator relation
\eqn\relnpa{
V_{n+1}^{\alpha_1}(z)V_m^{\alpha_2}(w)
= V_{n}^{\alpha_1}(z)V^{\alpha_2}_{m+1}(w)
}
Using the ring element $a_+$ instead of $a_-$ in \rcorr, we get other
operator relations. As before, we use the independence of the
position of the ring element in the correlation function.
Using the second equation of \eringb\ and \ind, we obtain
\eqn\corrw{\eqalign{
&\sum_{i=3}^N \langle  c{\bar c}V_{m_1}^{\gamma_i}(0)
c{\bar c}V_{n_2}^{\alpha_2}(1)c{\bar c}V_{n_3}^\alpha(\infty)
\prod_{j=3,j\neq i}^{N} \int V_{n_j}^{\alpha_j}
\rangle \cr
=&\sum_{i=3}^N\langle c{\bar c}V_{n_1}^{\alpha_1}(0)
c{\bar c}V_{m_2}^{\delta_i}(1)
c{\bar c}V_{n_3}^\alpha(\infty)
\prod_{j=3,j\neq i}^{N} \int V_{n_j}^{\alpha_j}\rangle\quad, \cr
}}
where $\gamma_i=(\alpha_1+\alpha_i+1)~{\rm mod}~p$,
$\delta_i=(\alpha_2+\alpha_i+1)~{\rm mod}~p$, $m_1=n_1+n_i-1+a_i$,
$m_2=n_2+n_i-1+b_i$,
$a_i\equiv{{(\alpha_1+\alpha_i+1)-\gamma}\over p}$ and
$b_i\equiv{{(\alpha_2+\alpha_i+1)-\delta}\over p}$.
Since the DK states chosen in \rcorr\ were arbitrary(modulo the
charge conservation condition), the above relation is valid for all
charge conserving correlation functions. This implies that in order that the
equality be always maintained, each term in the RHS of  \corrw\
should be equal to the corresponding term in the LHS. This leads to the
following operator relation (on focussing on the $i$-$th$ term
in \corrw.)
\eqn\relng{
V^{\gamma_i} _{m_1}(0)V^{\alpha_2}_{n_2}(1)=
V^{\alpha_1}_{n_1}(0)V^{\delta_i} _{m_2}(1)\quad,
}
Notice that when  $\delta_i,\gamma_i =(p-1)$ then we have the
``wrong edge" momenta appearing , then we use the relations recursively
to obtain relations between physical states. These relations involve
three operators and are of the form
\eqn\trel{
V^{p-2}_n(x) V^{1}_m(y) V^{1}_k(z) =
V^{0}_{n+1}(x) V^{0}_m(y)V^{0}_k(z)
}

Using the operator recursion relations \relnpa, \relng\ and \trel,
 we can reduce the correlation functions to
\eqn\pcorrg{\eqalign{
&\langle \langle  \prod_{i=1}^{N} \int  V_{n_i}^{\alpha_i} \rangle \rangle =
{\mu ^S }\Gamma (-S)
\langle c{\bar c}V_{0}^{0}(0)c{\bar c}V_{0}^{0}(1)~c{\bar c}
 V_{n3}^{p-2}(\infty)
(\int V_2^0)^{X} {1\over{R!}} (Q_-)^R \rangle \quad\cr
&={\mu ^S }\Gamma (-S) {\cal N}
\prod _{i=1}^X \prod _{j=1}^R \int d^2z_i d^2w_j
 {| z_i| }^{4\over p}
{| 1-z_i| }^{4\over p}
\prod_{i<j} {| z_i-z_j| }^{4(p+1)\over p}\cr
&\quad \times {| w_i| }^{-2\over (p+1)}
{| 1-w_i| }^{-2\over (p+1)} \prod_{i<j}{| w_i -w_j| }^{4p\over p+1}
\prod_{i,j}{| z_i-w_j| }^{-4}\cr
&={\mu ^S }\Gamma (-S) {\cal N} ({ X!})(\pi)^{X+R} (\rho)^{4RX}
\Delta (1-\rho')^X \Delta (1-\rho)^R\cr
&\prod_{i=1}^X \Delta (i\rho' -R) \prod_{i=1}^R\Delta (i\rho)\cr
&\prod_{i=0}^{X-1} \Delta (1-R+{2\over p}+i\rho')
\quad \Delta (1-R+{2\over p}+i\rho')
\quad \Delta (-1+R-{4\over p}-(X-1+i)\rho')\cr
&\prod_{i=0}^{X-1} \Delta (1-{2\over p+1}+i\rho)
\quad \Delta (1-{2\over p+1}+i\rho)
\quad \Delta (-1+2X-{4\over p+1}-(R-1+i)\rho)
 }}
where $\Delta(x) ={\Gamma(x)\over\Gamma(1-x)}$, $X =L+S-3$,
$\rho={p\over p+1}$, $\rho'={p+1\over p}$ and
${\cal N }$ is the normalisation factor associated with the operators.
 This can be further simplified to obtain
\eqn\zcorr{\eqalign{
\langle \langle \prod_{i=1}^L V_{n_i}^{\alpha_i} \rangle\rangle
&=\rho ~\mu^S ~{\Gamma(L+S-2)\over \Gamma (S+1)}\cr
&=\rho ~\mu^S~
{\Gamma( {{\sum_i n_i}\over 2}+ {{(\sum_i\alpha_i)+2}\over {2p}})
\over {\Gamma{(S+1)}}}\quad,
}}
where $S=\sum_{i}\left({{ n_i}\over 2}+{{\alpha_i}\over{2p}}\right)
-L +2+ {1\over p}$ and $\rho={p\over{p+1}}$.

Having obtained the correlation functions for integer values of S,
the results can be analytically continued to include non integer
values of S . However one must take care to preserve the $Z_2$
selection rule originating from the requirement of charge conservation
in the matter sector. The $Z_2$ selection rule implies that all correlation
functions that are $Z_2$ odd cannot be screened by the matter screening
operators and so they vanish. Note the $Z_2$ charge of the operator
$V_n^\alpha$ is $(-1)^{n+\alpha}$. In appendix A, we shall indicate
explicitly how the correlation functions can be obtained for the Ising
Model.

The result in \zcorr\ is consistent with the known results from matrix
model calculations.  For the Ising model, the result is in agreement
with the $\langle \epsilon^n \rangle$ and
$\langle \sigma\sigma\epsilon^n\rangle$ computed by Dijkgraaf and
Witten\matrixb. As a further proof of the validity of \zcorr, we
show in the sequel that this result is compatible with the multicritical
behaviour seen in the KdV/matrix models.

For later use, we state the results for the $(1,p)$ models.
\eqn\eonep{\eqalign{
\langle \langle \prod_{i=1}^L V_{n_i}^{\alpha_i} \rangle\rangle
&=\rho~\mu^S ~{\Gamma(L+S-2)\over \Gamma (S+1)}\cr
&=\rho~\mu^S~
{\Gamma( {{\sum_i n_i}\over 2}+ {{(\sum_i\alpha_i)+2}\over {2p}})
\over {\Gamma{(S+1)}}}\quad,
}}
where $S=\sum_{i}\left({{ n_i}}+{{\alpha_i}\over{p}}\right)
-L +2+ {2\over p}$ and $\rho={1\over{p}}$.

\newsec{Perturbed theories and Multicritical behaviour}
    Having derived all the genus zero correlation functions we can now
check to see if we can reproduce the multicritical behaviour seen in the
Matrix models. Distler\JDTG\ has shown that the $(2k-1,2)$ models can be
obtained by perturbations of the $(1,2)$ models, we will first reproduce
those results in this formalism, after which we look at the $(1,p)$ models
and show that the minimal models $(p+1,p)$ can be obtained from these models
by an appropriate perturbation.
Consider the perturbation of the action by the operator$V_1$ of the
form
\eqn\epfunc{
{\cal Z}_{pert}(\lambda_1) = \int  {\cal D} \phi~ {\cal D} X~
{\cal D}b{\cal D}c ~
\displaystyle {e^{-S_M - S_L -S_{gh}+ \mu \int V_0 +\lambda _1  \int V_1 }}
}
We expand the
exponential as a power series in the coupling $\lambda _1$ and
evaluate each term in the series using the results of the earlier
sections. We then sum the
series to obtain
\eqn\epertfunc{
{\cal Z}_{pert}(\lambda_1) = (1- \lambda_1)^{-1}{\cal Z} \quad,
 }
where ${\cal Z}$ is the unperturbed partition function as defined in
\pfunc.
Now if we set $\lambda_1=0$, the partition function diverges
indicating the presence of new multicritical behaviour as was noticed
in \JDTG.

 Now consider a perturbation of the $(1,2)$ model by the operator $V_k$,
in this model the operator $V_1$ is marginal and adding it to the action
only rescales all the correlation functions.
 The perturbed partition function is given by:
\eqn\epfunca{
{\cal Z}_{pert}(\lambda_1, \lambda_k) = \int  {\cal D} \phi~ {\cal D}X~
{\cal D}b{\cal D}c ~
\displaystyle {e^{-S_M - S_L -S_{gh}+
\mu \int V_0 +\lambda _1  \int V_1 +\lambda_k \int V_k}}}

   The correlation functions in the perturbed theory are
\eqn\epcorr{
\langle \langle \prod_{i=1}^{l} \int  V_{n_i} \rangle \rangle_{pert}
=\langle\langle exp^ {\lambda _1  \int V_1 +\lambda_k \int V_k}
\prod_{i=1}^{l} \int  V_{n_i} \rangle\rangle }
   where the $\langle\langle\cdots\rangle\rangle_{pert}$ stands for the
perturbed correlation function shown explicitly in the expression
on the R.H.S of \epcorr. (Note that the expression
$\langle\langle\cdots\rangle\rangle$ stands for the screened
correlation function as defined in  \eqcorr. )
Now we expand the the exponential interaction as a
series in the couplings to obtain

\eqn\epcorra{\eqalign{
\langle \langle  \prod_{i=1}^{l} \int  V_{n_i}& \rangle \rangle_{pert} =
\sum_m \sum_n {\mu^s \lambda_1^n \lambda_k^m \over m!~n!}\Gamma (-s)\cr
&\times\langle c{\bar c}V_{n1}(0)c{\bar c}V_{n2}(1)~c{\bar
c}V_{n3}(\infty) \prod_{i=3}^{l} \int  V_{n_i}
(\int V_1)^n (\int V_k)^m (\int V_0)^s{1\over{R!}} (Q_-)^R \rangle \quad\cr
}}

The Liouville charge conservation relation is
\eqn\econsp{
\sum_{i=1}^l n_i +n +km -(l+n+m+s) = - 3
}
which gives us the value for $s= \sum_{i=1}^l n_i + (k-1)m -l +3 $,
and $L+s-3 = \sum_{i=1}^l n_i +n +km $ where $L = l+m+n$
is the total number of operators in the correlation function. For ease
of notation, we shall use $\tilde{n}\equiv\sum_i n_i$ for the rest of
the section.

Using the computation for the correlation function in section 3, we find that
that the series in \epcorra\ can be rewritten as
\eqn \epcorrb{
\langle \langle \prod_{i=1}^{l} \int  V_{n_i} \rangle \rangle _{pert}=
\rho \sum_m \sum_n \mu^s {\lambda_1^n \lambda_k^m \over m!~n!~
(\tilde{n} + (k-1)m -l +3) !} (\tilde{n} +n +km )!}

Now we can do the summation over the index n, to obtain
\eqn \epcorrc{
\langle \langle \prod_{i=1}^{l} \int  V_{n_i} \rangle \rangle _{pert}=
\rho ~\mu^{(\tilde{n} - l +3)} (\widehat \lambda_1)^{-(\tilde{n} + 1)}
\sum_m {(\mu^{(k-1)}\widehat \lambda_1^{-k}\lambda_k)^m ~
(\tilde{n} +km )!\over m!~(\tilde{n}+ (k-1)m -l +3) !}}
where $\widehat \lambda_1= (1- \lambda_1)$.
Now if we define
$$
z = (\mu^{(k-1)}\widehat \lambda_1^{-k}\lambda_k)
\left({k^k\over (k-1)^{(k-1)}}\right)
$$
and replace the factorials with $\Gamma$ functions we obtain
\eqn \epcorrd{\eqalign{
\langle \langle \prod_{i=1}^{l} \int  V_{n_i} \rangle \rangle_{pert}& =
\rho ~\mu^{(\tilde{n} - l +3)} (\widehat \lambda_1)^{-(\tilde{n} + 1)}\cr
&\times\sum_m {z^m\over m!} \left({k^k\over (k-1)^{(k-1)}}\right)^{-m}
{\Gamma(\tilde{n} +km +1)\over \Gamma(\tilde{n} + (k-1)m -l +4)}\quad.
\cr }}
  Now the summation can be expressed in terms of the generalised
hypergeometric functions  (using the results of Appendix B) as:
\eqn\epcorre{
\langle \langle \prod_{i=1}^{l} \int  V_{n_i} \rangle \rangle_{pert} =
\rho ~\mu^{(\tilde{n} - l +3)} (\widehat \lambda_1)^{-(\tilde{n} + 1)}
{\Gamma(\tilde{n}+1)\over \Gamma(\tilde{n} -l +4)}
{ _K} F_{K-1}(\alpha_i;\gamma_j;z)}
where$$\alpha_i ={\tilde{n} + 1 \over k}+ {(i-1)\over k}\quad
i=1\ldots k \quad , $$
and $$\gamma _j ={\tilde{n} -l +4\over (k-1)}+ {(j-1)\over (k-1)}
\quad j=1\ldots (k-1)\quad .$$
We can extract the behaviour of the hypergeometric function in
limit $|z|\rightarrow \infty $
(using the results of Appendix B) to obtain:
\eqn\epcorrf{
\langle \langle \prod_{i=1}^{l} \int  V_{n_i} \rangle \rangle _{pert}=
\rho ~k^{-1}~\mu^{({{1 +\tilde n } \over k} - l +2)}
{}~\lambda_k^{({\tilde n  + 1\over k})}~
{\Gamma({{\tilde n  + 1}\over k})\over \Gamma({{{\tilde n +1}\over k} - l
+3)})}}
where $\tilde n =\sum_i n_i$.
 This is the expression for the correlation functions of the operators
$\langle \prod_{i=1}^{l} \int  V_{n_i} \rangle$ of the $(2k-1,2)$ model.
As we have just seen they can all be obtained by perturbation
from the (1,2) model , just as we can obtain the correlation functions
of  the k-th multicritical point in the one matrix model case.

   Now we will consider the (1,p) models and study their behaviour under
perturbation by physical operators, we will show that the minimal models
can be obtained from these models under perturbation.
The physical operators in these models are given by
$$
V_n^\alpha = \Delta(n + {{(\alpha+1)}\over p})exp{{[p(n-1)+\alpha]\phi
- [p(n-1)+\alpha+2]iX } \over {2\sqrt{p}}}
$$
where $\alpha=0,\ldots,(p-2)$.
 The correlation functions in these models can be obtained using the
ring to provide recursion relations, as was done for the minimal models.

Consider a perturbation of the $(1,p)$ model by the operator $V_k^0$
and $V_1^0$.
(In this model adding the operator $V_1^0$  to the action
 rescales all the correlation functions. This can be shown by a
calculation similar to the one in \epertfunc.)
 The perturbed partition function is given by:
\eqn\ppfunc{
{\cal Z}_{pert} = \int {\cal D} \phi~ {\cal D} X~ {\cal D}b{\cal D}c ~
\displaystyle {e^{-S_M - S_L -S_{gh}+
\mu \int V_0^0 +\lambda _1  \int V_1^0 +\lambda_k \int V_k^0}}
}

    The correlation functions in the perturbed theory are
\eqn\pcorr{
\langle \langle \prod_{i=1}^{l} \int  V_{n_i}^{\alpha_i} \rangle \rangle
_{pert}=\langle\langle exp^ {\lambda _1  \int V_1^0 +\lambda_k \int V_k^0}
\prod_{i=1}^{l} \int  V_{n_i}^{\alpha_i} \rangle \rangle
}
   where the $\langle\langle \cdots\rangle\rangle_{pert}$ stands for
the perturbed correlation function
shown explicitly in the expression on the R.H.S of \pcorr .
 (Note that the expression
$\langle\langle\cdots\rangle\rangle$ stands for the screened
correlation function . )

Now we expand the the exponential interaction as a
series in the couplings to obtain

\eqn\pcorra{\eqalign{
\langle \langle  \prod_{i=1}^{l}& \int  V_{n_i}^{\alpha_i} \rangle \rangle
 _{pert}= \sum_m \sum_n {\mu^s \lambda_1^n \lambda_k^m \over m! n! }
\Gamma(-s)\cr \times
&\langle c{\bar c}V_{n1}^{\alpha_1}(0)c{\bar c}V_{n2}^{\alpha_2}(1)~c{\bar c}
V_{n3}^{\alpha_3}(\infty) \prod_{i=3}^{l} \int  V_{n_i}^{\alpha_i}
(\int V_1^0)^n (\int V_k^0)^m (\int V_0^0)^s{1\over{R!}} (Q_-)^R \rangle
 \quad\cr }}

The Liouville charge conservation relation is
\eqn\consp{
p\tilde{n}+\tilde\alpha + (k-1)mp -(lp+sp) = -(2p+1)
}
which gives us the value for
 $$s= \tilde{n} + (k-1)m -l +2+{\tilde\alpha+2\over p}$$
and $$L+s-3 = \tilde{n} +n +km-1+{\tilde\alpha+2\over p}\quad,$$
where $L = l+m+n$ is the total number of operators in the correlation function
and $\tilde\alpha=\sum_i\alpha_i$.
For ease of notation, we shall use $\tilde\alpha\equiv\sum_i \alpha_i$ for
the rest of the section.

Using \eonep, we find that
that the series in \pcorra\ can be rewritten as
\eqn \pcorrb{
\langle \langle \prod_{i=1}^{l} \int  V_{n_i}^{\alpha_i}  \rangle \rangle
_{pert}= \rho \sum_m \sum_n \mu^s {\lambda_1^n \lambda_k^m \over m!~n!}~
{(\tilde{n} +n +km-1+{\tilde\alpha+2\over p}) !
\over (\tilde{n} + (k-1)m -l +2+{\tilde\alpha+2\over p})!}}

Now we can do the summation over the index $n$, to obtain
\eqn \pcorrc{\eqalign{
\langle \langle \prod_{i=1}^{l} \int  V_{n_i}^{\alpha_i}  \rangle \rangle
_{pert}& = \rho~\mu^{(\tilde{n} -l +2+{\tilde\alpha+2\over p} )}~
 (\widehat \lambda_1)^{-(\tilde{n} +{\tilde\alpha+2\over p})}\cr
&\times\sum_m {(\mu^{(k-1)}\widehat \lambda_1^{-k}\lambda_k)^m~
(\tilde{n}+km-1+{\tilde\alpha+2\over p} )!\over m!~
 (\tilde{n} + (k-1)m -l +2+{\tilde\alpha+2\over p}) !}\cr}}
where $\widehat \lambda_1= (1- \lambda_1)$.
Now if we define
$$z = (\mu^{(k-1)}\widehat \lambda_1^{-k}\lambda_k)
\left({k^k\over (k-1)^{(k-1)}}\right)$$
and replace the factorials with $\Gamma$ functions we obtain
\eqn \pcorrd{\eqalign{
\langle \langle \prod_{i=1}^{l} \int  V_{n_i} \rangle \rangle_{pert}& =
\rho ~\mu^{(\tilde{n} -l +2+{\tilde\alpha+2\over p} )}~
 (\widehat \lambda_1)^{-(\tilde{n} +{\tilde\alpha+2\over p})}\cr
&\times\sum_m {z^m\over m!} \left({k^k\over (k-1)^{(k-1)}}\right)^{-m}
{\Gamma(\tilde{n}+km+{\tilde\alpha+2\over p}) \over
 \Gamma(\tilde{n} + (k-1)m -l +3+{\tilde\alpha+2\over p})}
\cr }}
  Now the summation can be expressed in terms of the generalised
hypergeometric functions  (using the results of Appendix B):
\eqn\pcorre{\eqalign{
\langle \langle \prod_{i=1}^{l} \int  V_{n_i} \rangle \rangle _{pert}&=
\rho ~\mu^{(\tilde{n} -l +2+{\tilde\alpha+2\over p} )}~
 (\widehat \lambda_1)^{-(\tilde{n} +{\tilde\alpha+2\over p})} \cr
&\times{\Gamma(\tilde{n}+{\tilde\alpha+2\over p})
\over \Gamma(\tilde{n} -l +3+{\tilde\alpha+2\over p})}
{ _K} F_{K-1}(a_i;b_j;z)\cr}}
where $$a_i ={(\tilde{n}+{\tilde\alpha+2\over p})\over k}+ {(i-1)\over k}
\quad i=1\ldots (k-1)$$
 and
$$b_j ={(\tilde{n} -l +3+{\tilde\alpha+2\over p})\over(k-1)}
       + {(j-1)\over (k-1)}\quad j=1\ldots (k-1)$$
Notice that in the limit of $z\rightarrow \infty $ the correlation
functions diverge indicating the presence of new multicritical
behaviour. We can extract the behaviour of the hypergeometric function in this
limit  (using the results of Appendix B) to obtain:
\eqn\pcorrf{
\langle \langle \prod_{i=1}^{l} \int  V_{n_i} \rangle \rangle _{pert}=
{}~\rho~ k^{-1}~\mu^{({\tilde n p +{\tilde\alpha+2}\over pk}-l+2)}
{}~\lambda_k^{-(\tilde n +{\tilde\alpha+2 \over p})}~
{\Gamma({\tilde n p+{\tilde\alpha+2}\over pk})
\over \Gamma({\tilde n p +{\tilde\alpha+2}\over pk}-l+3)}\quad,
}
where $\tilde n=\sum_i n_i$ and $\tilde\alpha=\sum_i \alpha_i$.
Now if we choose the perturbing operator to be $V_2^0$ i.e, $k=2$ then
we obtain the minimal model (p+1,p) correlation functions as given in
eqn. \zcorr. For arbitrary $k$, one obtains the $(pk-p+1,p)$ models.

\newsec{Conclusions}
  In this paper we have shown that the correlation functions of $c<1$
string theory can be obtained using the operator identities provided by
the action of the ring inside correlation functions. We have also shown
that the multicritical behaviour of the matrix models can be reproduced
by perturbing the action by physical vertex operator states.
 and expanding in powers of the coupling constant.
In particular, we have shown that the $(2k-1,2)$ models can
be obtained as the k-th multicritical point of the $(1,2)$ models, and
the minimal models $(p+1,p)$ can be obtained as the second
multicritical point of the $(1,p)$ model.

The techniques introduced in this paper are sufficiently general and
can be used to evaluate the correlation functions in other models
like the $W_n$ string which have a similar ring structure.

The analytic continuation in the number of operators is relevant for
all computations for strings in non-trivial backgrounds which do not
have global translation invariance. However,
for such generic models it is not clear if
the ring structure exists and if it is sufficient to determine all the
correlation functions.

Although  it has been possible to obtain the correspondence between the
matrix models and the continuum picture completely for genus zero and
for the genus one partition function  the higher genus correlation
functions are required to ascertain the equivalence of the two
formulations, and remains  a challenging problem.
\bigskip
\noindent{\bf Acknowledgements:} We would like to thank P. Durganandini and P.
Majumdar for useful discussions. This work was done while one of us (S. G.)
was a Post-Doctoral Fellow at the Institute of Mathematical Sciences,
Madras.
\appendix{A}{Correlation Functions In the Ising Model}
   In this appendix we will indicate explicitly how the correlation functions
can be obtained for the Ising model.
We shall now use the ring elements to calculate arbitrary
N-point functions explicitly for the case of the Ising model which is
the  $(4,3)$ model. In the Ising model, $V_0^0$
corresponds to the identity operator, $V_0^1=\sigma$ and
$V_1^1=\epsilon$ correspond to the spin and energy operators
respectively(with appropriate Liouville dressing). $V_2^0$ is the
``physical'' screening operator.
One can prove the following
shift recursion relation. The arguments are identical to the one used
in deriving eqn.\relna .
\eqn\erecurb{
V_{n+1}^\alpha(z)V_m^\beta(w)
= V_{n}^\alpha(z)V_{m+1}^\beta(w)}
Now consider a charge conserving correlation function with one $a_+$
and DK states.
\eqn\ccorr{
\langle a_+(w) c{\bar c}V_{n}^0(0) c{\bar c}V_{m}^1(1)
c{\bar c}V_{r}^\alpha(\infty)
\prod_{i=1}^{L} \int V_{n_i}^0
\prod_{j=1}^{M} \int V_{m_i}^1
 {1\over{R!}} (Q_-)^R\rangle
}
Using the $w$ independence of the above correlation function, we
equate the value of the correlator at $w=0$ and $1$.
This gives us after using the second equation in \eringb
\eqn\acorrw{\eqalign{
&\sum_{k=1}^L \langle  c{\bar c}V_{n+n_k-1}^1(0) c{\bar c}V_{m}^1(1)
c{\bar c}V_{r}^\alpha(\infty)
\prod_{i=1,i\neq k}^{L} \int V_{n_i}^0
\prod_{j=1}^{M} \int V_{m_i}^1
\rangle \cr
&+
\sum_{k=1}^M\langle  c{\bar c}V_{n+m_k-1}^2(0) c{\bar c}V_{m}^1(1)
c{\bar c}V_{r}^\alpha(\infty)
\prod_{i=1}^{L} \int V_{n_i}^0
\prod_{j=1,j\neq k}^{M} \int V_{m_i}^1
\rangle \cr
&=\sum_{k=1}^{L}
\langle  c{\bar c}V_{n}^0(0) c{\bar c}V_{m+n_i-1}^2(1)
c{\bar c}V_{r}^\alpha(\infty)
\prod_{i=1,i\neq k}^{L} \int V_{n_i}^0
\prod_{j=1}^{M} \int V_{m_i}^1
\rangle \cr
&+ \sum_{k=1}^{M}
\langle  c{\bar c}V_{n}^0(0) c{\bar c}V_{m+m_k}^0(1)
c{\bar c}V_{r}^\alpha(\infty)
\prod_{i=1}^{L} \int V_{n_i}^0
\prod_{j=1,j\neq k}^{M} \int V_{m_i}^1
\rangle \cr
}}
Using eqn. \erecurb, one can see that every term inside each of the
sums are the same. One can now see that the following operator
relations provides a consistent solution to the above equality.
\eqn\erecurb{\eqalign{
V^2_m(z) V_n^1(w) &=V_m^0(z)V_{n+1}^0(w)\cr
V^2_m(z) V_n^0(w)  &= V_m^1(z)V_n^1(w)\cr
}}
However, $V^2_m$ belongs to the ``wrong-edge'' and is not
physical. However by using this relation twice, we obtain the following
recursion which involves only physical operators
\eqn\ewrecur{
V_m^1(z)V_n^1(w)V_r^1(t)=
V_{m+1}^0(z)V_n^0(w)V_r^0(t)\quad.
}
Equations \erecurb\ and
\ewrecur are sufficient to convert all charge conserving
correlation functions to those which are of the Dotsenko-Fateev type
and hence are computable. We shall now demonstrate this.
Consider the following correlation function
\eqn\eisinga{\eqalign{
\langle \langle \prod_{i=1}^L \int V_{n_i}^{\alpha_i} \rangle\rangle
= \mu^S \Gamma(-S)& \langle c{\bar c}V_{n_1}^{\alpha_1}(0)c{\bar c}
V_{n_2}^{\alpha_2}(1)~c{\bar c}V_{n_3}^{\alpha_3}(\infty)
\prod_{i=4}^L \int V_{n_i}^{\alpha_i} (\int V_0^0)^S
{1\over {R!}} (Q_-)^R \rangle \quad,
}}
where $S={{\sum_{i=1} n_i}\over 2} -L +2+ {{(\sum_{i=1}
\alpha_i)+2}\over 6}$
and $R=\sum_{i=1} ({{n_i}\over
2}+{{\alpha_i}\over 6}) +{{S}\over 3}$. When, $S$
and $R$ are positive integers, using  eqns. \erecurb\
and \ewrecur\ in eqn.\eisinga\ we obtain
\eqn\eisingb{
\langle \langle \prod_{i=1}^L \int V_{n_i}^{\alpha_i} \rangle\rangle
=\mu^S\Gamma(-S) \langle c{\bar c}V_0^0(0)c{\bar c}V_0^0(1)~c{\bar
c}V_1^1(\infty)
(\int V_2^0)^{(L+S-3)}{1\over{R!}} (Q_-)^R \rangle \quad.
}
The correlation function in the RHS can be explicitly computed using
the formula of Dotsenko and Fateev. We obtain
\eqn\esingc{
\langle \langle \prod_{i=1}^L \int V_{n_i}^{\alpha_i} \rangle\rangle
=\rho ~\mu^S ~{{(L+S-3)!} \over {S!}} \quad,
}
where $\rho={3\over4}$. This is in agreement with matrix model
results. For the cases
when $S$ and $R$ are not positive integers, the results are obtained
by analytic continuation. Of course, one has to take care that the
$Z_2$ invariance of the minimal models is not violated. One imposes
this by setting all non-$Z_2$ invariant correlators to zero by hand.
For the example of Ising model, the $Z_2$ charge of the operator
$V_n^\alpha$ is $(-1)^{n+\alpha}$. So the correlation function in
\eisinga is non-zero provided $\sum_i (n_i +\alpha_i)$ is
even. For such cases, the result one obtains after analytic
continuation in both $S$ and $R$ is
\eqn\acont{
\langle \langle \prod_{i=1}^L \int V_{n_i}^{\alpha_i} \rangle\rangle
=\rho ~\mu^S~
{{\Gamma( {{\sum_i n_i}\over 2}+ {{(\sum_i\alpha_i)+2}\over 6})}
\over {\Gamma{(S+1)}}}\quad.
}
To compare with the matrix model results in \matrixb, we  now exhibit
some results which follow form \acont
\eqn\ecompare{\eqalign{
\langle\langle\epsilon^n\rangle\rangle =\rho~ \left({\partial \over
\partial\mu}\right)^{n-3}\left(\mu^{{2n \over 3}-{2\over3}}\right)\quad,\cr
\langle\langle\sigma\sigma\epsilon^n\rangle\rangle =\rho~
\left({\partial\over \partial\mu} \right)^{n-1}
\left( \mu^{{2n \over 3}-{1\over3}}\right)\quad.
}}
This agrees with the result given in \matrixb\ upto trivial normalisation
factors. We have also checked that the results (for the above two sets
of correlation functions) in section 5, for the
$(3k-1,3)$ models agree with the two-matrix model results with $k$
labelling the various critical points.

\appendix {B}{Some relevant results on Generalised
Hypergeometric Functions}
   In this appendix we present the relevant details and the notations
used in the text to describe the generalised hypergeometric functions.
One can find more details in the following references \hyper .
The definition of the generalised hypergeometric function is
\eqn\defin{
_pF_q (\al_i, \gamma _i ;z) = {\gam\gamma{1} \gam\gamma{2}\ldots
\gam\gamma{q}\over \gam\alpha{1}
\gam\alpha{2}\ldots \gam\alpha{p}} \sum_{n=0} ^\infty{\Gamma(\alpha_1+n)\ldots
\Gamma(\alpha_p+n)\over \Gamma(\gamma_1+n)\ldots \Gamma(\gamma_q+n) }
{z^n\over n!}}
The domain of validity of the above expression is for $|z|<1$.
 There is also an identity satisfied by products of Gamma
functions which is
\eqn\pr{
\prod_{i=0} ^{p-1}\Gamma (\al +{i\over p})=(2\pi)^{(p-1)\over
2}(p)^{-p\al +1/2}\Gamma (p\al)}
   Using the product identity we can write
\eqn\hy{
_pF_q (\al_i, \gamma _i ;z) = {\Gamma(q\al)\over \Gamma(p\al)}\sum_{n=0}
^\infty  \left( {(q)^q\over (p)^p}\right)^n {z^n\over n!}
{\Gamma(p\al+np)\over \Gamma(q\gamma+nq)}}

There is also an  integral representation for the generalised
hypergeometric function
\eqn\rep{\eqalign{
_pF_q (\al_i, \gamma _i ;z) & = \left(\prod _{j=1}^p
{\gam\gamma{j}\over\Gamma(\al_j+1)\Gamma(\gamma_j-\al_{j+1})}\right. \cr
& \times\left.\int_0^1 \prod_{k=1}^p dt_k
(t_k)^{\al_{k+1}-1}(1-t_k)^{\gamma_j-\al_{j+1}-1}(1-zt_1 t_2\ldots
t_p)^{-\al_1}\quad+{\rm perms.} \right)\cr }}
 Using the integral representation, we extract the behaviour of the
hypergeometric function in the limit $|z| \rightarrow \infty$
\eqn\asymp{
\lim_{|z|\rightarrow \infty}{_pF_q (\al_i, \gamma _i ;z)} \sim
(-z)^{-\al _1} \prod _{j=1}^p
{\gam\gamma{j}\Gamma(\al_{j+1}-\al_1)
\over\Gamma(\al_j+1)\Gamma(\gamma_j-\al_{j+1})}\quad,}
where we choose $\alpha_1=min\{\alpha_i\}$.
We have used the results given above in deriving the
behaviour of the correlation
functions at the multicritical points.
\listrefs
\bye